\documentclass[prb,aps,twocolumn,superscriptaddress]{revtex4-1}
\usepackage{graphicx,color}
\usepackage{amsthm}
\usepackage{amsfonts}
\usepackage{algorithmic}
\usepackage{enumerate}
\usepackage{latexsym}
\usepackage{amsmath}
\usepackage{amssymb}
\usepackage{multirow}
\usepackage{array}
\usepackage{diagbox}
\usepackage[utf8]{inputenc}
\DeclareUnicodeCharacter{2032}{\ensuremath{'}}
\usepackage{subfigure}
\usepackage[colorlinks=true,citecolor=blue,linkcolor=blue]{hyperref}

\def\avg#1{\left\langle#1\right\rangle}

\emergencystretch=\maxdimen
\begin{document}
\title{Doping driven metal-insulator transition in disordered graphene}
\author{Kaiyi Guo}
\affiliation{Department of Physics, Beijing Normal University, Beijing 100875, China\\}
\author{Ying Liang}
\email{liang@bnu.edu.cn}
\affiliation{Department of Physics, Beijing Normal University, Beijing 100875, China\\}
\affiliation{Key Laboratory of Multiscale Spin Physics(Ministry of Education), Beijing Normal University, Beijing 100875, China\\}
\author{Tianxing Ma}
\email{txma@bnu.edu.cn}
\affiliation{Department of Physics, Beijing Normal University, Beijing 100875, China\\}
\affiliation{Key Laboratory of Multiscale Spin Physics(Ministry of Education), Beijing Normal University, Beijing 100875, China\\}

\begin{abstract}
Controlling the metal-insulator transition in graphene-based material is a crucial topic as it directly impacts its potential applications. Inspired by recent experiments, we study the effects of doping and bond disorder on metal-insulator transition in graphene within the Hubbard model on a honeycomb lattice. By using the determinant quantum Monte Carlo method, we first conduct tests on the value of $\avg{sign}$ under various parameters, such as electron density, on-site interactions, temperature, and lattice size, so as to select the appropriate parameters to alleviate the impact of the sign problem. Given the knowledge that bond disorder can lead to a mental-insulator transition, our study has revealed, after ruling out the influence of size effects, that the critical strength of disorder increases as the electron density decreases while decreasing as the on-site interactions increase. Furthermore, we compared our results with experimental data and concluded that, in actual graphene materials, the localization effect induced by doping plays a dominant role, resulting in an insulating phase.
\end{abstract}
\maketitle
\noindent
\section{Introduction}

Since the discovery of graphene, a honeycomb single layer of $sp^2$-bonded carbon atoms, it has attracted enormous attention because of its excellent electrical, structural, mechanical, and optical properties, which have always been the critical and challenging aspects of the research.\cite{doi:10.1126/science.1102896, Novoselov2005, Akinwande2019, Novoselov2012, doi:10.1126/sciadv.aaz4191} Due to its unique semimetal nature, intrinsic graphene can not provide sufficient conductivity for desired applications, and doping is considered as an optimal way to tailor the electronic structure of graphene,\cite{adfm.202203179,doi:10.1021/acsomega.2c06010} which allows for control of the Fermi level $E_F$  even pushes the van Hove singularity into the vicinity of $E_F$ and impact on superconducting pairing.\cite{PhysRevLett.125.176403,10.1063/1.3485059,PhysRevB.84.121410,PhysRevB.90.245114,PhysRevB.106.205144} Moreover, doping plays an extremely important role in various applications, such as photodetectors,\cite{Liu2022Graphene} sensors,\cite{kins2020} field-effect transistors,\cite{Meng2022Dual,Wang2019Free} and so on. In these applications, the regulation of metal-insulator transition (MIT) in graphene materials is very crucial, as it has a direct impact on further applications of these materials.\cite{Yang2023Unconventional,Li2021} Therefore, doping-dependent MIT in graphene is a worthwhile problem to investigate.

In essence, MIT can be driven by various mechanisms, resulting in different types of insulators: changing the chemical potential can produce a transition from a metal to a band insulator.\cite{PhysRevLett.83.2014,PhysRevB.92.155312} Strong correlations can drive metals into Mott insulators with an energy gap,\cite{Okamoto2004} while Anderson insulators originate from disorder-induced localized insulators, where no gap can be observed in the spectrum.\cite{PhysRevLett.102.136806} It is of great importance to tune and control MIT on graphene for applications.\cite{Osofsky2016,Ponomarenko2011,Yang2023Unconventional} However, the nature of the metal-insulator transition remains elusive despite tremendous effort due to the complex interaction of doping, chemistry, elastic strain, and other applied fields.\cite{Guzmán-Verri2019} There have been many experimental studies on MIT in graphene-based system. As early as 2009, researchers found that dosing atomic hydrogen on the surface of graphene would cause the system to transition from a metallic phase to an insulating phase and they discussed this phenomenon by possible transition to a strongly Anderson localized ground state.\cite{PhysRevLett.103.056404} Reports on MIT in nitrogen-doped and oxygen-doped graphene materials in 2016 further indicated that doping would transform the material from a metallic phase into an insulating phase.\cite{Osofsky2016} Recent reports also suggest the possibility of modulating MIT in graphene through an externally applied electric field.\cite{Yang2023Unconventional,Li2021}

Drawing inspiration from the aforementioned research, we conducted an investigation on the mechanical properties of graphene lattices at MIT. Due to the fact that doping leads to changes in carrier density and introduces disorder into the system at the same time,\cite{PhysRevLett.103.056404,Osofsky2016} while an applied electric field can also modulate electron density,\cite{Yang2023Unconventional,Li2021} we took into account both disorder and electron density in the system and studied their interplay and the impact they have on the MIT. In order to investigate strongly correlated problems with both disorder and doping,  the determinant quantum Monte Carlo (DQMC) method is a powerful tool\cite{PhysRevLett.83.4610,PhysRevLett.120.116601,PhysRevB.105.045132,PhysRevB.106.205149}.

In the context of QMC simulations, various interesting MIT phenomena have been reported in the honeycomb lattice.\cite{sorella1992semi,PhysRevX.6.011029,PhysRevB.97.045101}
For example, a disorder-induced nonmagnetic insulating phase is found to emerge from the zero-temperature quantum critical point, separating a semimetal from a Mott insulator at half filling.\cite{doi:10.1126/science.1204333} Furthermore, recent QMC simulations on a bilayer honeycomb lattice have identified a potential deconfined quantum critical point in interacting Dirac fermions as a new area of study for investigating the MIT.\cite{PhysRevLett.128.087201} Localization due to the on-site Coulomb interaction and disorder can also induce an insulating transition.\cite{PhysRevLett.120.116601}

In this paper, we completed our simulations by the DQMC method for cases with different electron densities and bond disorder strength to investigate the MIT in doped graphene with a disordered Hubbard model. Our main focus is on the impact of electron density, on-site Coulomb interaction, and bond disorder on the conductivity $\sigma_{dc}$. We analyzed the interplay between these three factors and found that doping increases conductivity, which is favorable for the formation of metallic phases, while disorder has the opposite effect. The impact of the on-site Coulomb interaction on $\sigma_{dc}$ depends on the particle-hole symmetry: at half-filling, the on-site Coulomb interaction suppresses conductivity, while deviating from half-filling can promote conductivity. Our study expands the understanding of MIT in honeycomb lattice through doping and disorder and may provide some inspiration for modulating MIT in experiments.

\noindent

\section{Model and methods}
The Hamiltonian for disordered Hubbard model on a honeycomb lattice is defined as
\begin{eqnarray}
\hat H=&&-\sum_{{\bf \avg{i,j}},\sigma}t_{\bf ij}(\hat c_{{\bf i}\sigma}^\dagger \hat c_{{\bf j} \sigma}^{\phantom{\dagger}}+\hat c_{{\bf j}\sigma}^\dagger \hat c_{{\bf i} \sigma}^{\phantom{\dagger}})-\mu \sum_{{\bf i}\sigma} \hat n_{{\bf i}\sigma}\notag\\
&&+U\sum_{{\bf i}}\hat n_{{\bf i}\uparrow}\hat n_{{\bf i}\downarrow}
\label{hamiltonian}
\end{eqnarray}
where $t_{\bf ij}$ represent the hopping amplitude between two nearest-neighbor sites ${\bf i}$ and ${\bf  j}$, $\hat c_{{\bf i}\sigma}^\dagger(\hat c_{{\bf j} \sigma}^{\phantom{\dagger}})$ is the creation (annihilation) operator of a spin-$\sigma$ electron at site ${\bf i}({\bf j})$, and $\hat n_{{\bf i}\sigma}=\hat c_{{\bf i}\sigma}^\dagger\hat c_{{\bf j} \sigma}^{\phantom{\dagger}}$ is the number operator, denotes the number of spin-$\sigma$ electrons at site ${\bf i}$. The chemical potential $\mu$ determines the density of the system, and when $\mu=\frac{U}{2}$, $n=1$, the system is half-filled, indicating the particle-hole symmetry. Here $U>0$ represent the on-site repulsive interaction.
Bond disorder is induced by modifying the matrix element $t_{\bf ij}$ of the hopping matrix, which is chosen from $t_{\bf ij}\in[t-\Delta/2,t+\Delta/2]$ and zero otherwise with a probability $P(t_{\bf ij})=1/\Delta$. We set $t=1$ as the energy scale. The strength of disorder can be characterized by $\Delta$, which represents the magnitude of the modification of matrix elements $t_{\bf ij}$ in the hopping matrix. In the presence of disorder, reliable results are obtained by taking an average of 20 disorder simulations, as it has been demonstrated to effectively avoids errors introduced by randomness.\cite{PhysRevLett.120.116601,PhysRevB.105.045132}

The DQMC method is employed to complete simulations on disordered Hubbard model of doped honeycomb lattice at finite temperature with periodic boundary condition. In DQMC, the partition function $Z=Tr e^{-\beta H}$ is represented as an integral over the configuration space of a set of interacting fermions on a lattice and the integral is completed by the Monte Carlo sampling. The imaginary time interval $(0,\beta)$ is discretely divided into $M$ slices of interval $\Delta \tau $, which is chosen as small as 0.1 to control the ``Trotter errors". The diagonalization of two-operator products can be achieved with simplicity; however, the same cannot be said for on-site interaction involving four-operator products as they need to be decoupled into quadratic terms before computation by a discrete Hubbard-Stratonovich (HS) field. Then, by analytically integrating the Hamiltonian quadratic term, the partition function can be converted into the product of two fermion determinants, where one is spin up and the other is spin down. The value of the fermion determinant is not always positive in calculations, except for a few exceptional cases, and this will cause sign problem. We calculated the average fermion sign $\avg{sign}$, which is the ratio of the integral of the product of up and down spin determinants to the integral of the absolute value of the product\cite{PhysRevB.92.045110}
\begin{eqnarray}
\label{sign}
\langle S \rangle &=
\frac
{\sum_{\cal X} \,\,
{\rm det} M_\uparrow({\cal X}) \,
{\rm det} M_\downarrow({\cal X})
}
{
\sum_{\cal X}  \,\,
| \, {\rm det} M_\uparrow({\cal X}) \,
{\rm det} M_\downarrow({\cal X}) \, |
}
\end{eqnarray}
to measure the severity of the sign problem. $\avg{sign}=1$ indicates the absence of sign problem.

To study the MIT of the system, we computed the $T$-dependent DC conductivity from calculating the momentum $\textbf{q-}$ and imaginary time $\tau$-dependent current-current correlation function
$\Lambda_{xx}(\textbf{q},\tau)$:
\begin{eqnarray}
\label{DC}
\sigma_{dc}(T)=\frac{\beta^2}{\pi}\Lambda_{xx}(\textbf{q}=0,\tau=\frac{\beta}{2})
\end{eqnarray}
where $\Lambda_{xx}(\textbf{q},\tau)$=$\left<\hat{j}_x(\textbf{q},\tau)\hat{j}_x(\textbf{-q},0)\right>$, $\beta$=$1/T$, $\hat{j}_x(\textbf{q},\tau)$ is the Fourier transform of time-dependent current operator $\hat{j}_x(\textbf{r},\tau)$ in the $x$ direction:
\begin{eqnarray}
\label{transform}
\hat{j}_x(\textbf{r},\tau) = e^{H\tau/h}\hat{j}_x(\textbf{r})e^{-H\tau/h}
\end{eqnarray}
where $\hat{j}_x(\textbf{r})$ is the electronic current density operator, defined in Eq.(\ref{J}).
\begin{eqnarray}
\label{J}
\hat{j}_x(\textbf{r}) = {i}\sum_{\sigma}t_{i+\hat{x},i}\times(c_{i+\hat{x},\sigma}^{+}c_{i\sigma}-c_{i \sigma}^{+}c_{i+\hat{x},\sigma})
\end{eqnarray}
Eq.(\ref{DC}) has been used for MIT in the Hubbard model in many studies.\cite{PhysRevLett.83.4610,PhysRevLett.87.146401, PhysRevLett.90.246401,PhysRevLett.98.046403,PhysRevLett.120.116601, PhysRevB.106.205149,PhysRevB.105.045132,zhang2021metal}

\noindent

\section{Results and discussion}

\begin{figure}[htbp]
\centering
\includegraphics[scale=0.45]{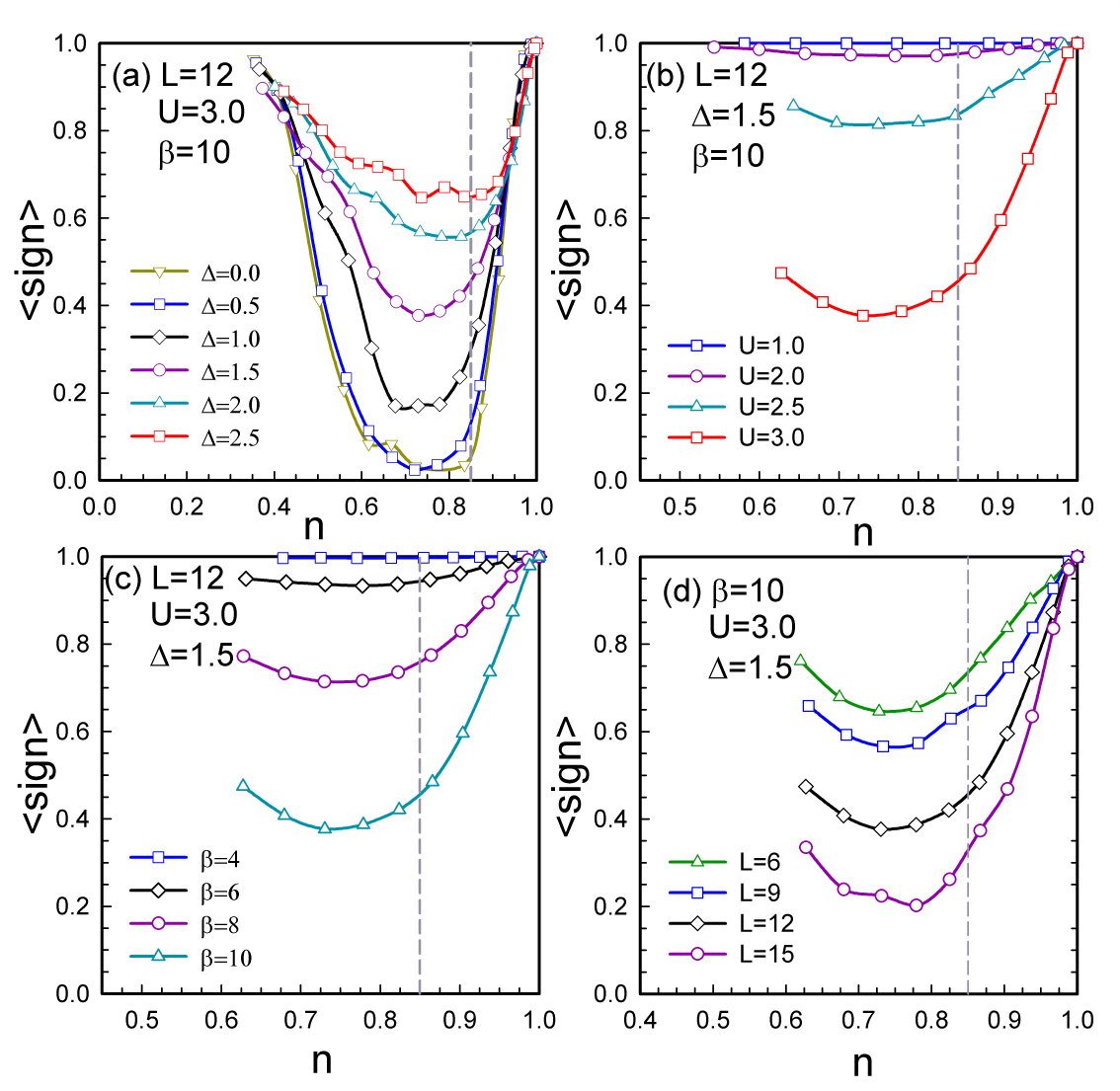}
\caption{$\avg{sign}$ as a function of the electron density $n$ for different values of (a) disorder, (b) on-site interaction, (c) temperature, and (d) lattice size. The dashed line indicates the case of $n=0.85$.}
\label{Fig:sign}
\end{figure}

As the system is doped away from half-filled, the particle-hole symmetry no longer exists, resulting in a sign problem. We have known that $\avg{sign}\sim e^{-\beta N_s\gamma}$, where $\gamma$ relies on the values of $n$ and $U$. In the case of a given fixed $n$ value, $\gamma$ is a monotonic function of $U$; whereas, with respect to a designated $U$ value, $\gamma$ is relatively small at certain specific values of $n$. To ensure the reliability of the data, the value of the average sign $\avg{sign}$, given by Eq.(\ref{sign}), was calculated and the corresponding results are presented in Fig.\ref{Fig:sign}. We present the average sign $\avg{sign}$ as a function of the electron density $n$ for different values of (a) disorder strength, (b) on-site interaction, (c) temperature, and (d) lattice size. Our studies were conducted in the region of $n\geq0.85$, with the dashed line indicating the case of $n=0.85$. Obviously, when the system is doped, the average sign deviates from $1$ and starts to decrease rapidly. The sign problem becomes more severe as the inverse temperature, interaction strength, lattice size increase, while introducing disorder can alleviate the sign problem to some extent. This is consistent with the preceding investigations.\cite{PhysRevB.92.045110}

Fig.\ref{Fig:sign}(a) shows the variation of average sign with respect to $n$ for different disorder strengths $\Delta$ at $L=12$, $U=3.0$ and $\beta=10$. It can be observed that in the clean limit, $\Delta=0.0$, the sign problem is severe and the calculation is almost impossible even with minor doping. However, the introduction of disorder partially alleviates the sign problem, and in the regime $\Delta\geq1.0$, which is of our primary interest, the sign problem is effectively suppressed. Fig.\ref{Fig:sign}(b) exhibits the influence of on-site interaction on the sign problem when $L=12$, $\Delta=1.5$ and $\beta=10$, implying that a larger $U$ greatly exacerbates the sign problem. Moreover, it is observable that when $U<2.5$, $\avg{sign}\sim 1$, making the impact of the sign problem almost negligible. The similar consequence is also evident in the Fig.\ref{Fig:sign}(c): when $\beta <6$, the sign problem has a minimal impact; however, as $\beta$ increases and the temperature decreases, the sign problem becomes increasingly severe. Fig.\ref{Fig:sign}(d) displays the effect of lattice size $L$ on the sign problem: as the lattice size increases, $\avg{sign}$ decreases and the sign problem becomes dire.

\begin{figure}[htb]
\centering
\includegraphics[scale=0.45]{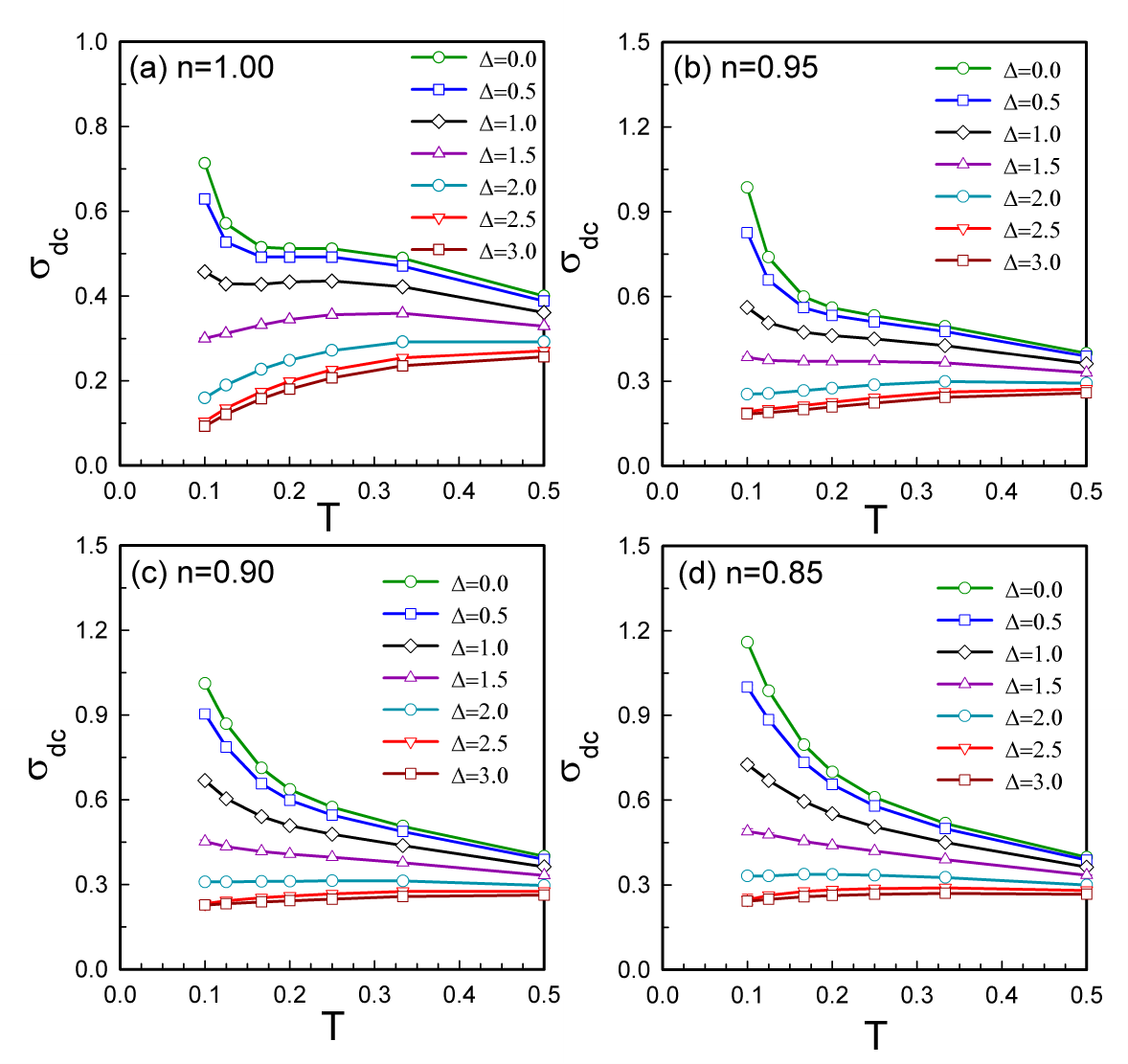}
\caption{DC conductivity $\sigma_{dc}$ as a function of temperature $T$ calculated on the $N=2\times L^2=288$ lattice with $U=2.0$ for various disorder strength $\Delta$. (a)-(d) represent different electron densities of $n=1.00,0.95,0.90,0.85$ respectively. }
\label{Fig:dcviaD}
\end{figure}

\begin{figure}[htbp]
\centering
\includegraphics[scale=0.45]{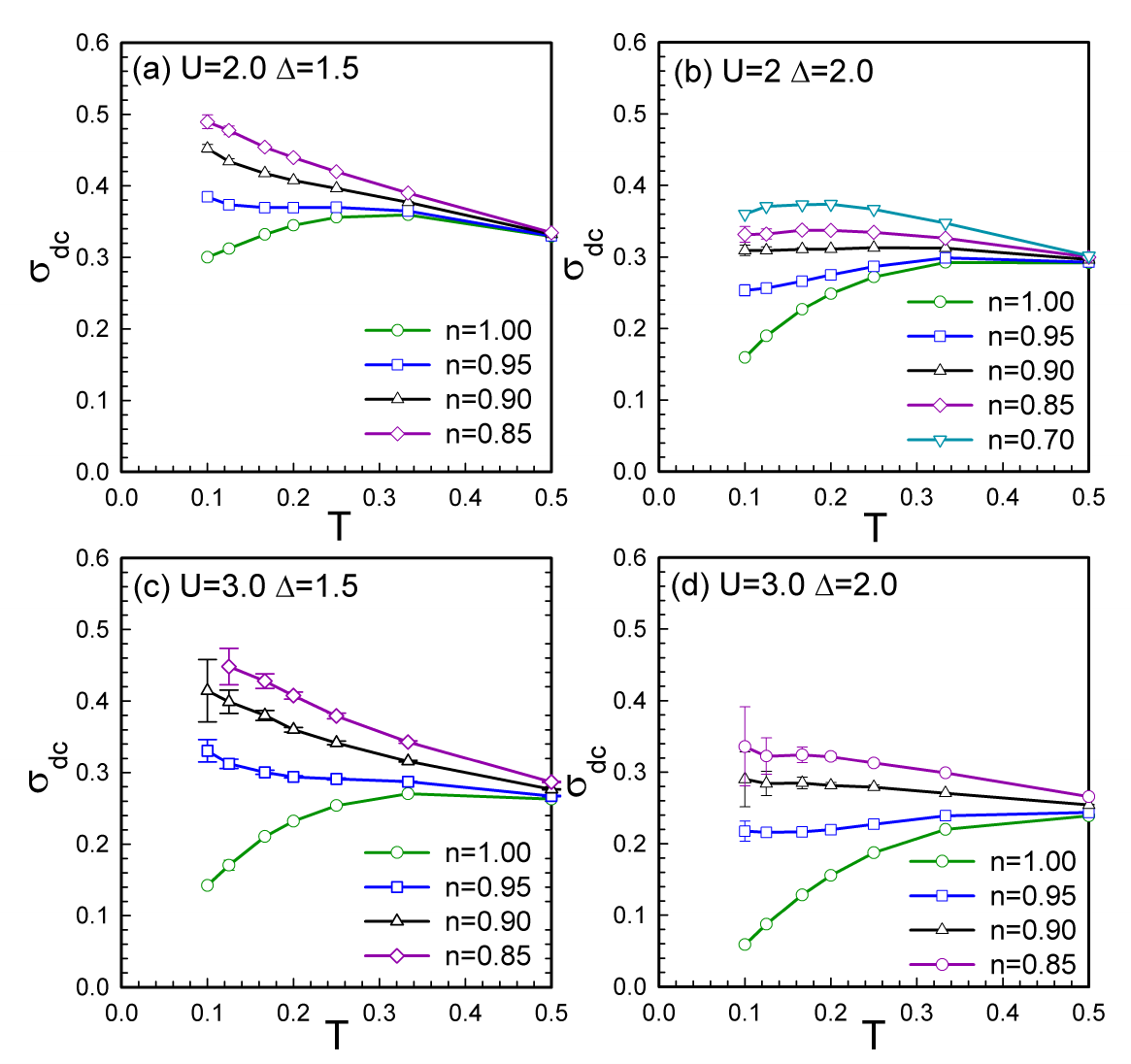}
\caption{ DC conductivity $\sigma_{dc}$ as a function of temperature $T$ calculated on the $N=2\times L^2=288$ lattice in the case of fixed disorder strength (a)(c)$\Delta=1.5$ and (b)(d)$\Delta=2.0$ under different value of electron density $n=1.00,0.95,0.90,0.85$. Top panel is about $U=2.0$ and lower panel is about $U=3.0$.}
\label{Fig:dcvian}
\end{figure}

Given the significance of the sign problem, along with the computational processing time considerations, we opt to utilize a lattice size of $L=12$ as the primary subject of inquiry in this article, building upon the conclusion presented in Fig.\ref{Fig:sign}. In Fig.\ref{Fig:dcviaD}, the dc conductivity is shown as a function of the temperature $T$ for several values of the disorder strength $\Delta$. The values are computed on the $L=12$ lattice with coupling strength $U=2.0$. Figs.\ref{Fig:dcviaD}(a)-(d) represent the situations under different density: (a) $n=1.00$; (b) $n=0.95$; (c) $n=0.90$; and (d) $n=0.85$. We have known that the system behaves as a mental in the clean limit at half-filling with the coupling strength $U=2.0$\cite{PhysRevLett.120.116601}, which means that in the low-temperature regime, $d\sigma_{dc}/dT<0$ and $\sigma_{dc}$ diverges as the temperature is further decreased to the limit $T\rightarrow0$. Then consider about the situations with bond disorder, the system will transfer from metallic to insulating phase, indicating by $d\sigma_{dc}/dT>0$ at low-$T$, with increasing value of $\Delta$, as is shown in Fig.\ref{Fig:dcviaD}(a). At this condition, the critical disorder strength for MIT $\Delta_c$ is currently between 1.5 and 2.0. When the system deviates from half-filling, as is shown in Figs.\ref{Fig:dcviaD}(b)-(d), distinct insulation behavior is only observed for $\Delta>1.5$. From this, we may draw the conclusion that in disordered systems, doping will increase the critical disorder strength $\Delta_c$ required for MIT. The impact of electron density $n$ on MIT will be further discussed in Fig.\ref{Fig:dcvian}.

\begin{figure}[htbp]
\centering
\includegraphics[scale=0.45]{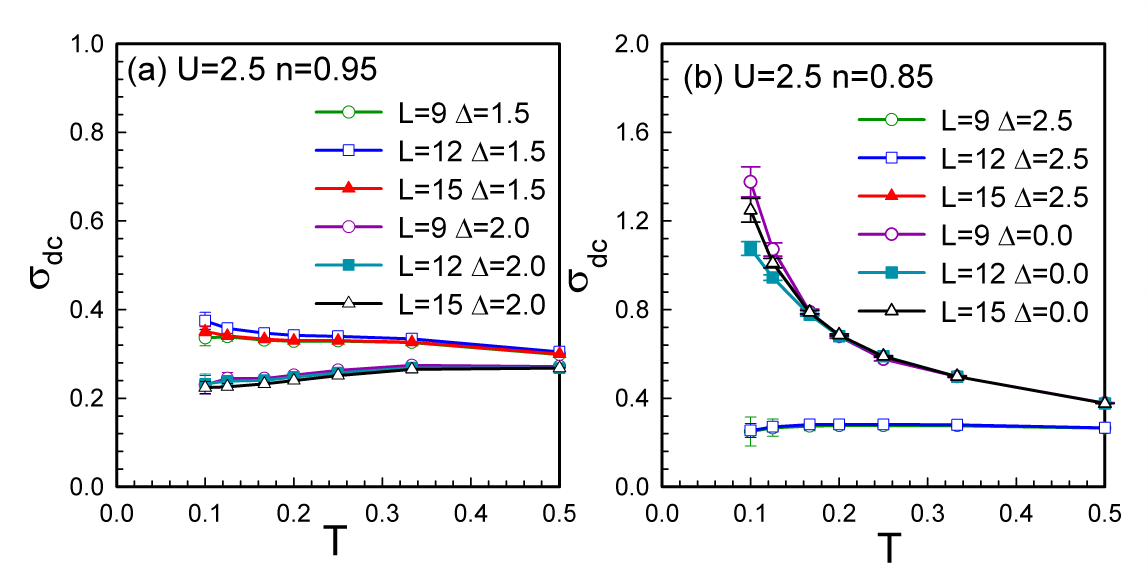}
\caption{ DC conductivity $\sigma_{dc}$ as a function of temperature T for different disorder strength $\Delta$ at $U=2.5$ for electron density (a)$n=0.95$ and (b)$n=0.85$. The lattice scaling is $L=9,12,15$ respectively. }
\label{Fig:scaling}
\end{figure}

To exclude the influence of system size being smaller than the localization length on insulation, we compute the finite-size effect. Fig.\ref{Fig:scaling} exhibits the response of the conductivity $\sigma$ to the lattice size $L=9,12,15$, with respect to different electron density (a) $n=0.95$, (b) $n=0.85$ and varying values of disorder (a) $\Delta=1.5$, $2.0$ and (b) $\Delta=0.0$, $2.5$. Upon comparison, it is evident that both the metallic and insulating phases are minimally affected by system size in terms of conductivity. Additionally, Fig.\ref{Fig:scaling}(a) illustrates that the critical disorder strength values remain consistent across varying lattice dimensions of $L=9,12,15$. As the computational simulation time rapidly increases with an increase in lattice size, and a larger $L$ suggests more severe sign problems while deviating from half-filling, it is reasonable that we selected $L=12$ as the primary focus of our study.

In Fig.\ref{Fig:dcvian}, we further investigate the impact of electron densities $n$ on the MIT. Fig.\ref{Fig:dcvian}(a) and Fig.\ref{Fig:dcvian}(b) respectively demonstrate the effect of $n$ on the $\sigma_{dc}-T$ curve for $L=12$, $U=2.0$, and the disorder strength (a)$\Delta=1.5$ and (b)$\Delta=2.0$: When $\Delta=1.5$, as shown in Fig.\ref{Fig:dcvian}(a), at $n=1.00$, the system exhibits an insulating phase due to hopping disorder, while deviating away from half-filling, the conductivity $\sigma_{dc}$ increases with decreasing temperature, indicating metallic behavior, thus demonstrating a MIT induced by doping; When $\Delta=2.0$, however, as shown in Fig.\ref{Fig:dcvian}(b), the system will always remain in an insulating phase irrespective of the variation in $n$. We have also included the $\sigma_{dc}-T$ curve for $n=0.7$, which reveals that within our measurement range, doping will not induce a MIT when the disorder strength $\Delta=2.0$. A similar situation can be observed at on-site Coulomb interaction $U=3.0$, as shown in Fig.\ref{Fig:dcvian}(c)$L=12$, $U=2.0$, $\Delta=1.5$ and (d)$L=12$, $U=2.0$, $\Delta=2.0$. Doping induces a transition from an insulating to a metallic phase at $\Delta=1.5$, whereas there is no metallic phase observed in the range of $n\leq0.85$ when $\Delta=2.0$.

\begin{figure}[htbp]
\centering
\includegraphics[scale=0.46]{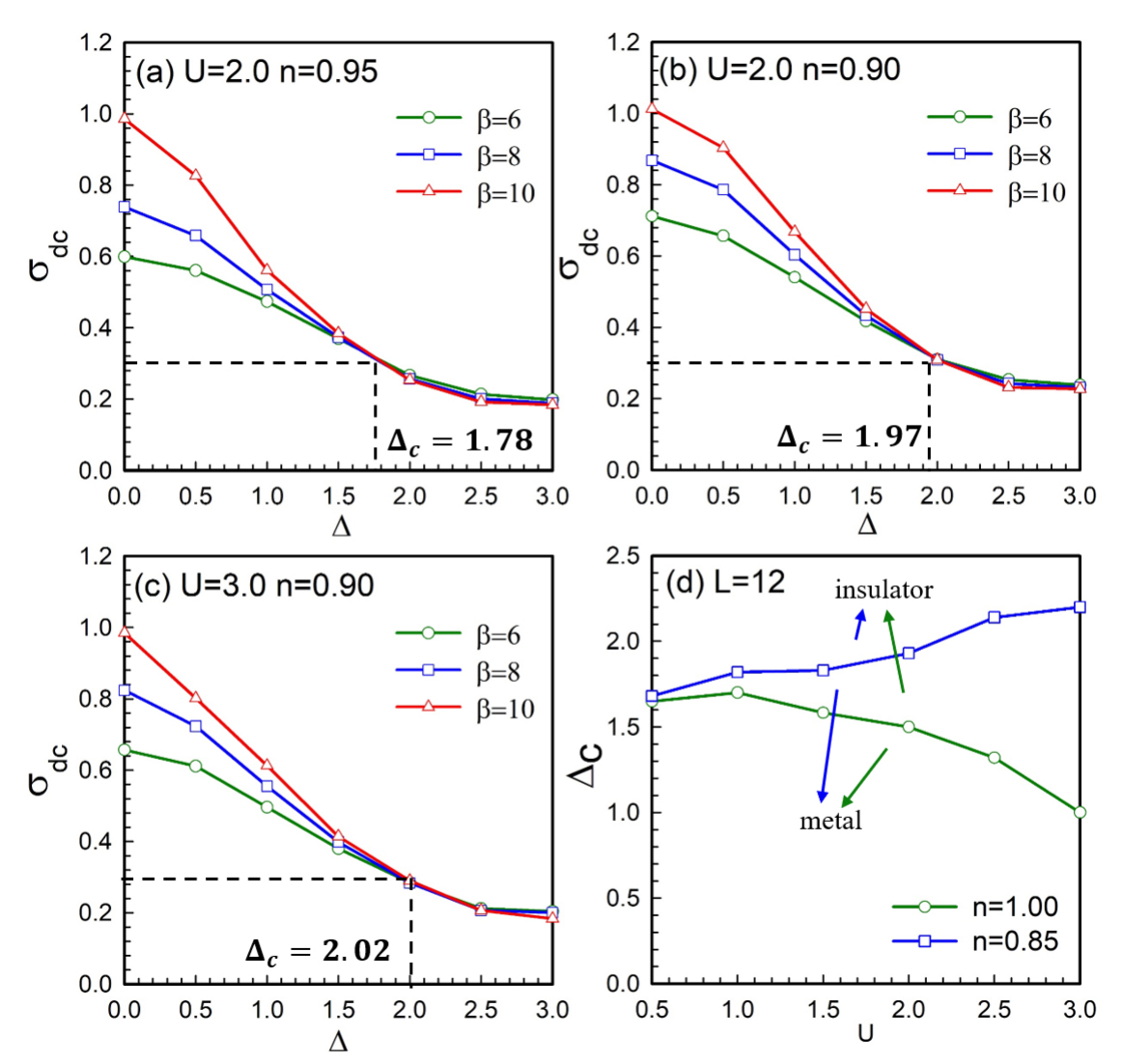}
\caption{ DC conductivity $\sigma_{dc}$ as a function of disorder strength $\Delta$ for three lowest temperature $\beta=6,8,10$. (a)$L=12, U=2.0, n=0.95$, (b)$L=12, U=2.0, n=0.95$ and (c)$L=12, U=3.0, n=0.90$. (d)critical disorder strength $\Delta_c$ as a function of $U$ at different $n$, segmenting the image into two parts: mental and insulator.}
\label{Fig:dc-D}
\end{figure}

To obtain a more accurate determination of the critical disorder strength for the MIT, we plot the variation of conductivity $\sigma_{dc}$ with disorder strength $\Delta$ at the three lowest temperatures $\beta=6,8,10$ in Fig.\ref{Fig:dc-D}(a)-(c). When $\Delta<\Delta_c$, the $\sigma_{dc}$ increases with decreasing temperature, exhibiting metallic behavior, while for $\Delta>\Delta_c$, the $\sigma_{dc}$ decreases with decreasing temperature, exhibiting insulating behavior. The three curves in each subplot of Fig.\ref{Fig:dc-D} intersect nicely at a point where the conductivity $\sigma_{dc}$ becomes temperature-independent, marking the critical point of MIT. Here, (a) corresponds to $L=12,U=2.0,n=0.95$; (b) corresponds to $L=12,U=2.0,n=90$; and (c) corresponds to $L=12,U=3.0,n=0.90$. We have conducted extensive calculations to obtain the values of $\Delta_c$ for different parameters and plot the variation of $\Delta_c$ with on-site Coulomb interaction $U$ for electron density $n=1.00$ and $n=0.85$ in Fig.\ref{Fig:dc-D}(d), where the curves above denote the insulating phase and the curves below denote the metallic phase. An interesting phenomenon can be observed: as $n=1.00$ and the system is half-filled, the critical disorder strength $\Delta_c$ of MIT decreases with an increase in $U$, indicating a suppressing effect of $U$ on the metallic state; whereas when $n=0.85$ and the system deviates from half-filling, $\Delta_c$ increases with an increase in $U$, signifying a promoting effect of $U$ on the metallic state.

\begin{figure}[htbp]
\centering
\includegraphics[scale=0.46]{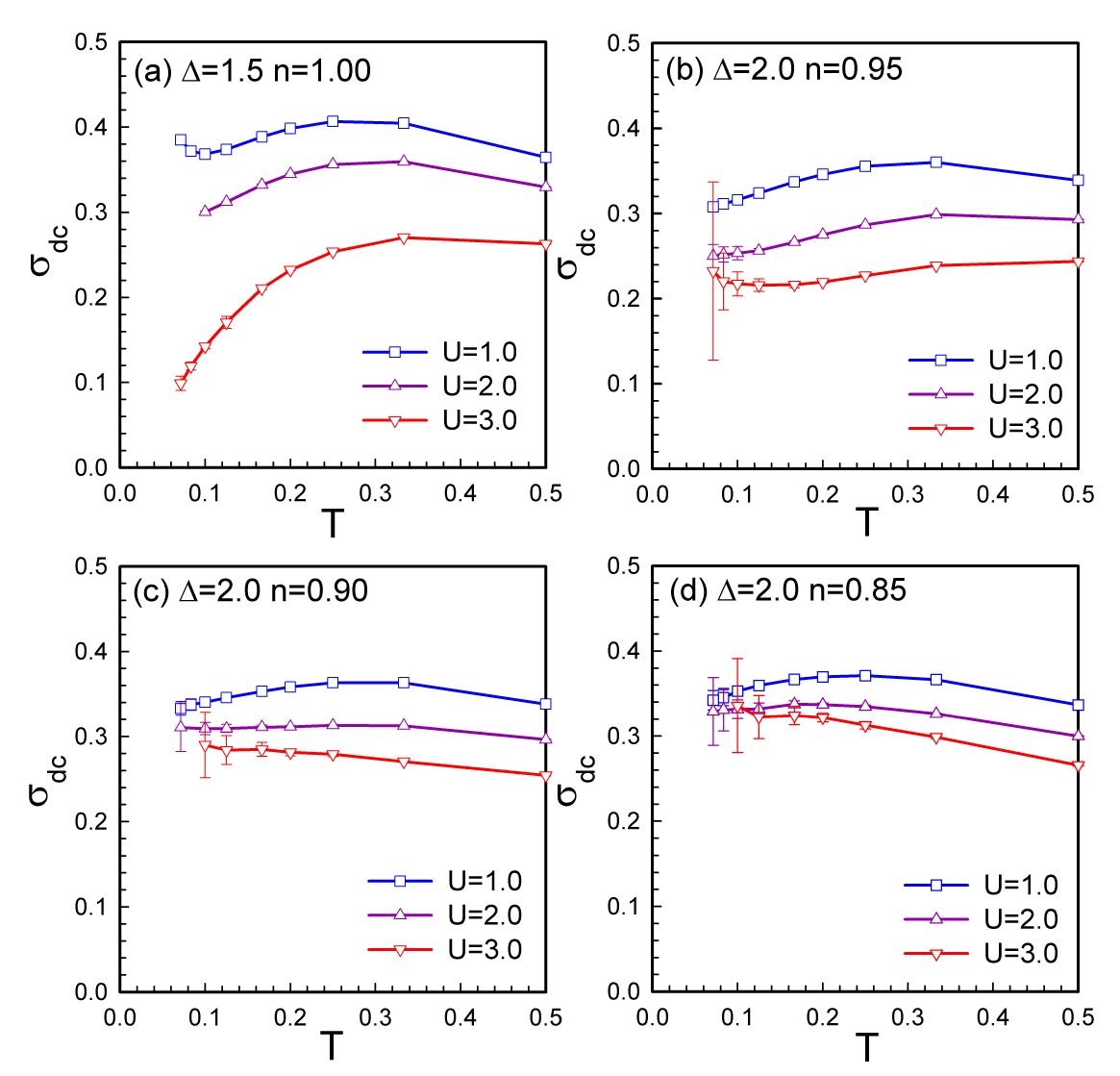}
\caption{ DC conductivity $\sigma_{dc}$ as a function of temperature T for different on-site interaction $U$ st density (a)$n=1.00$, (b)$n=0.95$, (c)$n=0.90$ and (d)$n=0.85$. Disorder strength $\Delta=1.5$ in (a) and $\Delta=2.0$ in (b)-(d).}
\label{Fig:dcviaU}
\end{figure}

Next we move on to the  role of $U$ in the MIT for half-filled and doped cases. Fig.\ref{Fig:dc-D}(d) demonstrates that at $n=1.0$ and $\Delta=1.5$, an increase in $U$ drives the system from a metallic state to an insulating state, whereas at $n=0.85$ and $\Delta=2.0$, an increase in $U$ leads the system from an insulating state to a metallic state. We set $n=1.00,0.95,0.90,0.85$ in Fig.\ref{Fig:dcviaU}(a)-(d). In order to observe the phase transition, we set the disorder strength to $\Delta=1.5$ for half-filling and $\Delta=2.0$ for deviations from half-filling, respectively. Furthermore, we set the minimum temperature parameter to $\beta=14$. Although this approach incurs a significant degree of error, it still yields valuable information. We then proceed to calculate the temperature dependence of the conductivity $\sigma$ at different on-site Coulomb interactions $U=1.0,2.0,3.0$. Fig.\ref{Fig:dcviaU}(a) shows the transition of the system from a metallic state to an insulating state as the on-site Coulomb interaction $U$ increasing, while Fig.\ref{Fig:dcviaU}(b)-(d) show the transition in the opposite direction. At $U=1.0,2.0$, the system shows insulating phases and at $U=3.0$, the system exhibits metallic phase. Overall, Fig.\ref{Fig:dcviaU} demonstrates that in half-filled systems, $U$ has a suppressing effect on the metallic state, while in doped systems, $U$ has a promoting effect on the metallic state.

\noindent

\section{Conclusion}

In summary, we employed the determinant quantum Monte Carlo method to investigate the regulatory effects of doping and disorder on the metal-insulator transition process in graphene materials. We discussed the factors affecting the MIT, including doping, temperature, lattice size and on-site Coulomb interactions by carrying out calculations for variations of the DC conductivity $\sigma_{dc}$ with temperature under different values, utilizing the reciprocal of the variation of $\sigma_{dc}$ with temperature $T$ to determine the metallic or insulating phase of the system. Through our calculations, we have reached the conclusion that doping increases conductivity and induces a transition from insulator to metal phase, while disorder has the opposite effect.

In experiments, substitutional doping or adsorbate doping often simultaneously alters the carrier density and introduces disorder, thus making the competition between doping and disorder important in the study of MIT in graphene materials. Our calculations show that when doping and disorder coexist, a larger disorder strength may cause the system to transition from the metal phase to the insulating phase. This finding is consistent with the metal-insulator transition phenomenon observed in hydrogen, nitrogen, and oxygen substitutional doped graphene materials in experiments.\cite{PhysRevLett.103.056404,Osofsky2016} Our research contributes to a deeper understanding of the mechanisms underlying the metal-insulator transition in graphene materials, and may be helpful in the development of applications for graphene materials.


\noindent
\section{Acknowledgements}

This work was supported by NSFC (No. 11974049).
The numerical simulations in this work were performed at HSCC of
Beijing Normal University.

\bibliography{reference}
\end{document}